\documentclass{llncs}

\usepackage{times}
\usepackage{color}
\usepackage{hyperref}
\usepackage{graphics}
\usepackage{wrapfig}

\title{PredatorHP Attacks Interval-Sized Regions\\
{\large (Competition Contribution)\vspace*{-4mm}
\thanks{The work was supported by the Czech
Science Foundation project 17-12465S, the IT4IXS: IT4Innovations Excellence in 
Science project (LQ1602), and the internal BUT project FIT-S-17-4014}
}}

\author{Michal Kotoun \and Petr Peringer \and Veronika
\v{S}okov\'{a}\thanks{Jury member.} \and Tom\'{a}\v{s} Vojnar\vspace*{-1.5mm}}

\institute{Brno University of Technology, FIT, IT4Innovations Centre of
Excellence, Czech Republic\vspace*{-5.5mm}}

\begin{document}

\maketitle

\begin{abstract} This paper describes shortly the basic principles of the
PredatorHP (Predator Hunting Party) shape analyzer and presents its recent
improvements. One of the most visible changes is the way PredatorHP handles
interval-sized memory regions, which is particularly useful for dealing with
arrays whose size is not fixed in advance. Further, the paper characterizes
PredatorHP's participation in SV-COMP 2019, pointing out its strengths and
weakness and the way they were influenced by the latest changes in the
tool.\end{abstract}

\vspace*{-6mm}\section{Verification Approach}\vspace*{-1mm}

We first briefly recap the main ideas of the verification approach behind
Predator and then discuss improvements that have been done in the latest version
of the tool.

\vspace*{-2mm}\subsection{The Predator Shape Analyzer}\vspace*{-1mm}

Predator is implemented as a GCC plug-in on top of the Code Listener framework
\cite{eurocast11}.
In particular, the Code Listener framework transforms the Gimple code produced
by GCC as an intermediate representation of the input program into a bit more
concise representation over which Predator runs.

The main aim of Predator is \emph{shape analysis} of sequential C programs that
use low-level C pointer statements to implement various kinds of lists (singly-
or doubly-linked, possibly circular, nested, and/or shared). 
Predator looks for various kinds of \emph{memory-related errors} (invalid
pointer dereferences, double free operations, memory leaks, etc.), and it also
checks validity of \emph{assertions} present in the code.

\enlargethispage{5mm}

Predator uses \emph{abstract interpretation} based primarily on the domain of
\emph{symbolic memory graphs} (SMGs).
SMGs are oriented graphs with two main kinds of nodes and two kinds of edges.
In particular, the nodes can be classified as \emph{objects} and \emph{values}.
Objects represent either particular \emph{regions} of memory allocated by a
single allocation statement or sets of (unboundedly many) memory regions linked
into different kinds of \emph{list segments} (which are automatically recognised
by the analyser).
Regions are marked as \emph{valid} or \emph{invalid} (i.e., deallocated).
The latter are kept until they are pointed in some way.
Values are stored in objects and classified as pointers or other data values.
Edges are classified as \emph{has-value} and \emph{points-to} edges.
The has-value edges start at a certain offset of memory regions, and points-to
edges point to target regions again with some offset.
%
%
More details on the abstract domain of SMGs can be found in~\cite{sas13}.

The primary abstraction used in Predator summarizes uninterrupted sequences of
singly- or doubly-linked memory regions into appropriate kinds of list segments.
Apart from that, Predator can abstract numerical values (either values stored in
regions, offsets, or sizes of regions) using intervals with constant bounds.
The constants used as the bounds have a pre-defined maximum/minimum value
defined in the configuration of Predator (+32/-32 for SV-COMP'19).
If the maximum/minimum value is exceeded, the bound is set to plus or minus
infinity.

Predator uses \emph{summaries} to speed up analysis of programs structured into
functions.
Analysis of recursive programs is, however, not supported (or, more precisely,
they are handled up to some configurable maximum depth).

\begin{wrapfigure}[11]{r}{77mm}
  \centering
  \vspace*{-8mm}
    \includegraphics{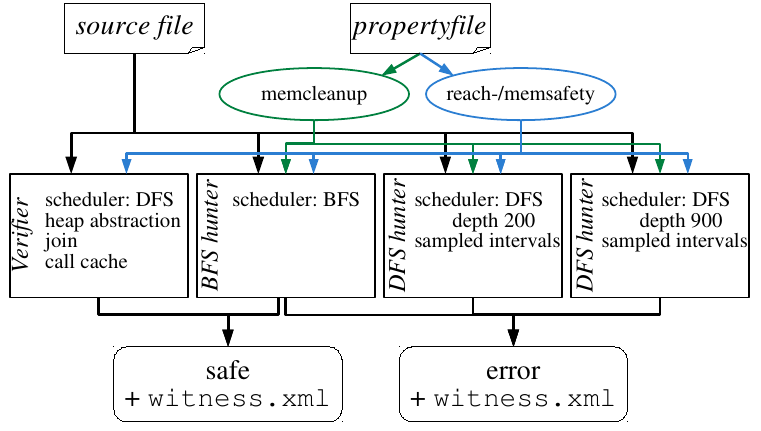}
\end{wrapfigure}

The so-called \emph{Predator Hunting Party} (PredatorHP)
\cite{svcomp15}, whose flow of control is shown on the right, is used
to increase both the efficiency as well as precision of the analysis.
Namely, PredatorHP starts the base Predator analyser---called as the
\emph{Predator verifier}---in parallel with several \emph{Predator hunters},
which are restricted versions of the analyser not allowed to use the
list-segment abstraction, to join SMGs that are not semantically equal, nor to
use function summaries with call parameter matching based on SMG entailment.

While the Predator verifier can claim a program correct, it cannot report errors
(since these could be false alarms due to the use of abstraction).
Predator hunters are classified as \emph{breadth-first} (BFS) and
\emph{depth-first} (DFS).
We use two Predator DFS hunters that can go at most 200 or 900 Code-Listener
instructions deep in the execution, and one Predator BFS hunter for which the
number of instructions to be performed is not limited, but it is limited
implicitly by the time available for the verification.
Usually, the hunters can only report errors, but they cannot claim a program
correct since they do not use memory abstraction, and they perform bounded
analysis of dynamic data structures only.
The only exception is the case when the verified program is finite-state, and
the entire state space gets explored by the BFS hunter in the given time limit.

As soon as some of the Predator instances reports a result which it is
authorised to announce (i.e., the verifier can report correctness while the
hunters announce errors---with the above mentioned exception for finite-state
programs), the other Predator instances get killed.
Checks whether some of the Predators has already produced a useful result are
done by the top-level script of PredatorHP with a pre-defined frequency.

\enlargethispage{6mm}

\vspace*{-2mm}\subsection{Recent Modifications of PredatorHP}\vspace*{-1mm}

The main issue that we tackled in the latest version of Predator is working with
\emph{interval-sized memory regions}, which arise when allocating structures or
arrays of parametric size.
Despite even older versions of Predator were able to create such regions, the
way in which they could have been treated in the subsequent analysis of the
program was very limited.
In particular, it was impossible to dereference interval-sized regions, and
hence Predator was very weak when analysing programs with structures or arrays
of an in-advance-not-fixed size.
This situation has been improved in the latest version of Predator in the
following pragmatic way.

Namely, whenever the current version of Predator hits a conditional statement
that would previously yield an interval value with fixed bounds (e.g., executing
the statement \texttt{if (n
>=0 \&\& n<10)} where \texttt{n} is originally unconstrained), it will split
the further run of the analysis into as many branches as the number of values in
the interval is, each of them evaluating for a concrete value from the interval.
After the split, no further interval-based allocations and dereferences, which
the previous version of Predator used to fail on, happen.
Though this solution is rather simple, it works nicely in many cases.
Of course, it can lead to a memory explosion in the case when the intervals are
large, but then the analysis fails with no answer as it used to fail before.

The above modification of Predator concerned dealing with memory regions whose
size is given by an interval with finite bounds.
In case one of the bounds is infinite, Predator has been extended to
\emph{sample} the interval and perform the further analysis with the sampled
values.
Currently, the sampling is done simply by taking some number of concrete values
from the given interval starting/ending with the bound that is fixed (of course,
for memory regions, unboundedness from above does only make sense).
The number of considered samples is currently set to 3.
Of course, this strategy cannot be used to soundly verify correctness of
programs, and so it is used for detecting bugs only.

Another issue that was resolved in the latest version of PredatorHP is detection
of \emph{invalid dereferences} of \emph{objects local to a block} from outside
of the block.
For that, a support of the \emph{clobber} instruction of Gimple was added.
This instruction is used in Gimple to terminate the life time of local variables
of code blocks, and it was previously not supported by Predator.
Now, whenever the instruction is encountered, the concerned memory region is
marked as deallocated, and further dereferences of that region are detected as
erroneous.

As another improvement, we have strengthened the way Predator uses to check
whether two pointers point to \emph{different addresses}.
So far, roughly speaking, inequality of pointers could only have been
established for pointers pointing to different valid objects, null and non-null
objects, the same object with different offsets, or to different ends of a
doubly-linked list segment with at least two elements.
We have now added one more way how inequality can be established, namely, when
comparing pointers to a valid region and to an invalid region where the invalid
one was allocated later than the valid one (which we can check due to the way
the objects are numbered in Predator).
Indeed, in such a case, both of the regions must have existed at the same time,
the valid object is continuously valid since then, and the two objects must lie
on different addresses since the address of a continuously valid object could
not have been recycled.

Next, we have added a support for checking whether \emph{all dynamically
allocated memory has been deallocated} when a function with the
\texttt{noreturn} attribute (such as \texttt{abort} or \texttt{exit}) is called.
This modification is, in fact, quite simple---we just need to check whether the
SMG representing the memory in such a moment does not contain any valid
dynamically allocated object.


\enlargethispage{5mm}

\vspace*{-2mm}\section{Strengths and Weaknesses}\vspace*{-1mm}

The main strength of PredatorHP is that it treats manipulation with various
kinds of unbounded lists in a \emph{sound} and \emph{efficient} way.
Predator hunters then allow it to quickly handle programs with a small finite
state space (e.g., benchmarks on locks) and avoids many false alarms that could
otherwise happen.

The main weakness of Predator has traditionally been its weak treatment of
non-pointer data.
We have tried to improve on that using the described heuristics for dealing with
intervals of integers with a specific aim to improve the way Predator handles
arrays of parametric size.
The results of Predator on SV-COMP'19 benchmarks with arrays show that the
heuristic did indeed help.
Moreover, they also helped on some benchmarks with arithmetic operations on data
fields of lists.
On the other hand, the modified treatment of intervals caused, somewhat
paradoxically, Predator some losses as well.
Namely, in the heap reachability category, it removed some unknown results
caused previously by dereferences via interval-based offsets, but Predator then
produced false alarms due to other imprecisions in handling non-pointer data.

Due to the added support of \texttt{clobber} instructions, Predator detects
invalid memory accesses in new benchmarks accessing variables outside of the
block in which they were created.
All other new heuristics described above did also help in some cases.

Another weakness of Predator is that it is specialized in dealing with lists,
and hence it does not handle structures such as trees or skip-lists (that is, it
handles them but in a~bounded way only).

\section{Tool Setup and Configuration}

The source code of PredatorHP used in SV-COMP'19 is freely available on the
Internet\footnote{\fontsize{8.5}{8.5}\selectfont\href{http://www.fit.vutbr.cz/research/groups/verifit/tools/predator-hp}{\texttt{http://www.fit.vutbr.cz/research/groups/verifit/tools/predator-hp}}}.
The file \texttt{README-SVCOMP-2019} shipped with the source code describes how
to build the tool. To run it, the script \texttt{predatorHP.py} can be invoked.
%
%
The script does not impose any resource limits other than terminating its child
processes when they are no longer needed. 
%
%
In SV-COMP'19, PredatorHP participated in the \textit{MemSafety} category and in
the \textit{HeapReach} sub-category of \textit{ReachSafety} category.
More information about the setting of PredatorHP used in SV-COMP'19 can be found
here:
\href{http://sv-comp.sosy-lab.org/2019/systems.php}{\texttt{http://sv-comp.sosy-lab.org/2019/systems.php}}.

\vspace*{-2mm}\section{Software Architecture, Project, and
Contributors}\vspace*{-1mm}

Predator is implemented in C++ with a use of Boost libraries as a GCC plug-in
based on the Code Listener framework \cite{eurocast11}. PredatorHP is
implemented as a Python script. Predator is an open source software project
distributed under the GNU General Public License version 3. The main author of
Predator is Kamil Dudka. Besides him and the PredatorHP team, Petr M\"uller, and
numerous other people listed in the \texttt{docs/THANKS} file in the 
distribution of Predator have contributed to the distribution of Predator.



\begin{thebibliography}{10}

\bibitem{sas13}
K.~Dudka, P.~Peringer, and T.~Vojnar.
\newblock Byte-Precise Verification of~Low-Level List Manipulation.
\newblock In {\em Proc. of SAS'13}, {\em LNCS} 7935, pp. 214--237, Springer, 2013.

\bibitem{eurocast11}
K. Dudka, P. Peringer, and T. Vojnar.
\newblock An Easy to Use Infrastructure for Building Static Analysis Tools.
\newblock In {\em Proc. of EUROCAST'11}, {\em LNCS} 6927, pp. 527--534,
2012.


\bibitem{svcomp15}
P. M\"{u}ller, P. Peringer, and T. Vojnar. 
\newblock Predator Hunting Party (Competition Contribution). 
\newblock In {\em Proc. of TACAS'15}, {\em LNCS 9035}, pp. 443--446, 2015. 

\end{thebibliography}
\end{document}